\documentclass[conference]{IEEEtran}
\usepackage[utf8]{inputenc}
\usepackage[english]{babel}
\usepackage{graphicx}
\usepackage{xcolor}
\usepackage{cite}
\usepackage[T1]{fontenc} 
\usepackage{amssymb}
\usepackage{amsfonts}
\usepackage{amsmath}
\usepackage{multirow}
\usepackage{soul}
\usepackage{mathtools}
\usepackage[acronym]{glossaries}
\usepackage{amsmath}
\usepackage{multirow}
\usepackage{algorithm}
\usepackage{algorithmic}
\usepackage{wrapfig}
\def\BibTeX{{\rm B\kern-.05em{\sc i\kern-.025em b}\kern-.08em
    T\kern-.1667em\lower.7ex\hbox{E}\kern-.125emX}}
\makeatletter
\def\blfootnote{\xdef\@thefnmark{}\@footnotetext}
\makeatother
\usepackage{lipsum}

\usepackage{caption}

\DeclareMathOperator{\EX}{\mathbb{E}}

\DeclareMathOperator*{\amax}{arg\,max}
\newcommand*{\starnr}{\stepcounter{equation}\tag{\theequation}}


\begin{document}


\newacronym{iot}{IoT}{Internet of Things}
\newacronym{pb}{PB}{power beacon}
\newacronym{hap}{HAP}{hybrid access point}
\newacronym{ula}{ULA}{uniform linear array}
\newacronym{rf}{RF}{radio frequency}
\newacronym{ue}{UE}{user}
\newacronym{irsa}{IRSA}{irregular slotted aloha}
\newacronym{eh}{EH}{energy harvesting}
\newacronym{pmf}{PMF}{probability mass function}
\newacronym{los}{LOS}{line-of-sight}
\newacronym{nlos}{NLOS}{non-line-of-sight}
\newacronym{fcsi}{F-CSI}{full channel state information}
\newacronym{acsi}{A-CSI}{average channel state information}
\newacronym{mrt}{MRT}{maximum ratio transmission}
\newacronym{mrc}{MRC}{maximal ratio combining}
\newacronym{psd}{PSD}{power spectral density}
\newacronym{cdf}{CDF}{cumulative distribution function}
\newacronym{snr}{SNR}{signal-to-noise ratio}
\newacronym{mdp}{MDP}{markov decision process}
\newacronym{wpt}{WPT}{wireless power transfer}
\newacronym{bs}{BS}{base station}
\newacronym{tdma}{TDMA}{time division multiple access}
\newacronym{pdf}{PDF}{probability density functions}
\newacronym{rl}{RL}{reinforcement learning}
\newacronym{crdsa}{CRDSA}{contention resolution diversity slotted Aloha}
\title{Decentralized RL-Based Data Transmission Scheme for Energy Efficient Harvesting}

\author{\IEEEauthorblockN{Rafaela Scaciota$^{1}$, Glauber Brante$^{2}$, Richard Souza$^{3}$, Onel~Lopez$^{1}$, \\Septimia~Sarbu$^{1}$, Mehdi Bennis$^{1}$, and Sumudu~Samarakoon$^{1}$}
\IEEEauthorblockA{\textit{$^1$Centre for Wireless Communication, University of Oulu, Finland}}
\IEEEauthorblockA{\textit{$^2$ Federal University of Technology - Paran\'a, Brazil}}
\IEEEauthorblockA{\textit{$^3$Federal University of Santa Catarina, Brazil}}
}

\maketitle
\begin{abstract}
The evolving landscape of the \gls{iot} has given rise to a pressing need for an efficient communication scheme. As the \gls{iot} user ecosystem continues to expand, traditional communication protocols grapple with substantial challenges in meeting its burgeoning demands, including energy consumption, scalability, data management, and interference. In response to this, the integration of wireless power transfer and data transmission has emerged as a promising solution. This paper considers an \gls{eh}-oriented data transmission scheme, where a set of users are charged by their own multi-antenna \gls{pb} and subsequently transmits data to a \gls{bs} using an \gls{irsa} channel access protocol. We propose a closed-form expression to model energy consumption for the present scheme, employing \gls{acsi} beamforming in the wireless power channel. Subsequently, we employ the \gls{rl} methodology, wherein every user functions as an agent tasked with the goal of uncovering their most effective strategy for replicating transmissions. This strategy is devised while factoring in their energy constraints and the maximum number of packets they need to transmit. Our results underscore the viability of this solution, particularly when the \gls{pb} can be strategically positioned to ensure a strong \acrlong{los} connection with the user, highlighting the potential benefits of optimal deployment.
\end{abstract}


\section{Introduction}
\label{sec:intro}

\blfootnote{
This work has been partially supported in Brazil by CAPES, Finance Code 001, CNPq (402378/2021-0, 305021/2021-4, 307226/2021-2). This work is funded by the European Union Project CENTRIC under Grant Agreement (GA 101096379), VERGE (GA 101096034) the project Infotech R2D2, the Research Council of Finland (former Academy of Finland) (GA 348515), and the Finnish Foundation for Technology Promotion. Views and opinions expressed are however those 
\begin{wrapfigure}{l}{2cm} 
\vspace{-0.2cm}
\centering
\includegraphics[scale=1]{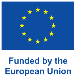}
\end{wrapfigure}
of the author(s) only and do not necessarily reflect those of the European Union or the European Commission (granting authority). Neither the European Union nor the granting authority can be held responsible for them. Corresponding author: rafaela.scaciotasimoesdasilva@oulu.fi.}

By the end of $2023$, it is projected that \acrfull{iot} users will constitute approximately $50\%$ of all networked users worldwide, as reported in~\cite{cisco}. In this dynamic and swiftly evolving realm of the \gls{iot}, the need for efficient and dependable communication methods has reached unprecedented importance~\cite{Nguyen.22}. As this expansive and continually growing ecosystem of \gls{iot} users continues to scale up, conventional communication protocols face substantial challenges in meeting the requirements of this burgeoning network. Within this context, the integration of wireless power transfer and wireless data transfer has emerged as a pivotal solution, opening up a new era of possibilities for \gls{iot} users~\cite{Choi.18}.

In this context, works like~\cite{Nguyen.22b, Vu.21} have demonstrated the feasibility of harvesting \gls{rf}-energy from sources such as a \acrfull{pb}. Wireless-powered systems face the challenge of efficiently utilizing harvested energy while minimizing packet collisions for users. To address this issue, simple Aloha protocols are commonly employed. These protocols are favored for their signaling simplicity and energy efficiency at the transmitter, which ultimately helps reduce collisions and optimize energy utilization. With this aim, prior works consider the use of \acrfull{irsa} protocol for powering the users over \gls{rf}. In~\cite{Chen.22}, the authors present a feedback-aided \gls{irsa} scheme that improves the user energy efficiency by optimizing the transmit power and the number of packet replicas using high-throughput transmission probability distributions. With the focus on improving wireless powered systems, the authors in~\cite{Demirhan.18} further extend the discussion by proposing an \gls{irsa}-based uncoordinated random access scheme for \gls{eh} nodes. It is considered a scheme where each user has a battery that is recharged by an \gls{eh} system. The results showcase optimized probability distributions for packet replicas and highlight the improvement of performance in \gls{irsa} protocol.  In~\cite{Silva.21}, the authors introduce an \gls{irsa} protocol tailored for resource-constrained nodes in wireless energy transfer environments. Therein, the concept of a \gls{hap} is used and the optimal threshold value that maximizes throughput in these unique networks is investigated.

A learning-based solution for addressing communication protocols integrated with wireless power transfer challenges can be found in~\cite{Li.22} where the authors delve into the integration of \gls{irsa} with \gls{rf} users, with a particular emphasis on the critical task of optimizing the number of packet replicas. What sets this research apart from previous approaches is the utilization of a \textit{Q}-learning-based methodology. This approach allows for the dynamic adjustment of the number of replicas based on the energy levels of the users, resulting in substantial improvements in the success rate of transmissions. However, it is worth noting that certain gaps related to the utilization of the \gls{irsa} protocol in wireless-powered systems, such as scalability, interference, latency, and hardware cost, still exist in the literature. These gaps represent opportunities for further research and exploration in this evolving field.

Inspired by the existing works, we present a novel joint data transmission and \gls{eh} for \gls{iot} networks. In this scheme, individual users are powered by their respective \gls{pb} as a distributed \gls{eh}, employing \acrfull{acsi} beamforming techniques. We compare the use of \gls{acsi} with \gls{fcsi} technique which assumes the availability of perfect instantaneous \gls{fcsi} for the user-\gls{pb} link. Subsequently, each user proceeds to transmit data to a central \gls{bs} using an \gls{irsa} channel access protocol. Our paper provides a comprehensive closed-form expression that accurately models energy consumption within this framework. From an \gls{eh} perspective, the energy beamforming scheme empowers a multi-antenna power beacon to efficiently deliver power to individual users, solely relying on the first-order statistics of channel conditions. This approach effectively mitigates interference from other user-\gls{pb} systems within the setup, thanks to the integration of distributed \gls{eh}.

 Moreover, to address the \gls{eh}-oriented data transmission scheme, we adopt an approach based on \acrfull{rl}. In this context, each user acts as an agent with the objective of discovering their optimal strategy for transmitting replicas, taking into consideration their energy limitations and the maximum number of packets to be transmitted. We leverage independent \textit{Q}-learning to showcase the scalability of our system. The scalability allows for an increase in the number of users, providing practical feasibility for addressing the challenges in \gls{eh}-oriented data transmission schemes. The significance of our learning methodology becomes evident when we draw comparisons with the \gls{crdsa} channel access protocol, which consistently sends the same number of packets per user~\cite{Meloni.12}. Our numerical results further emphasize the positive impact of incorporating the \textit{Q}-learning method, demonstrating about $18\%$ increase in successful transmissions per frame compared to a baseline scheme employing the \gls{crdsa} channel protocol without any learning process.

\section{System Model}
\label{sec:model}
\begin{figure}[t]
  \centering
  \includegraphics[width=8.8cm]{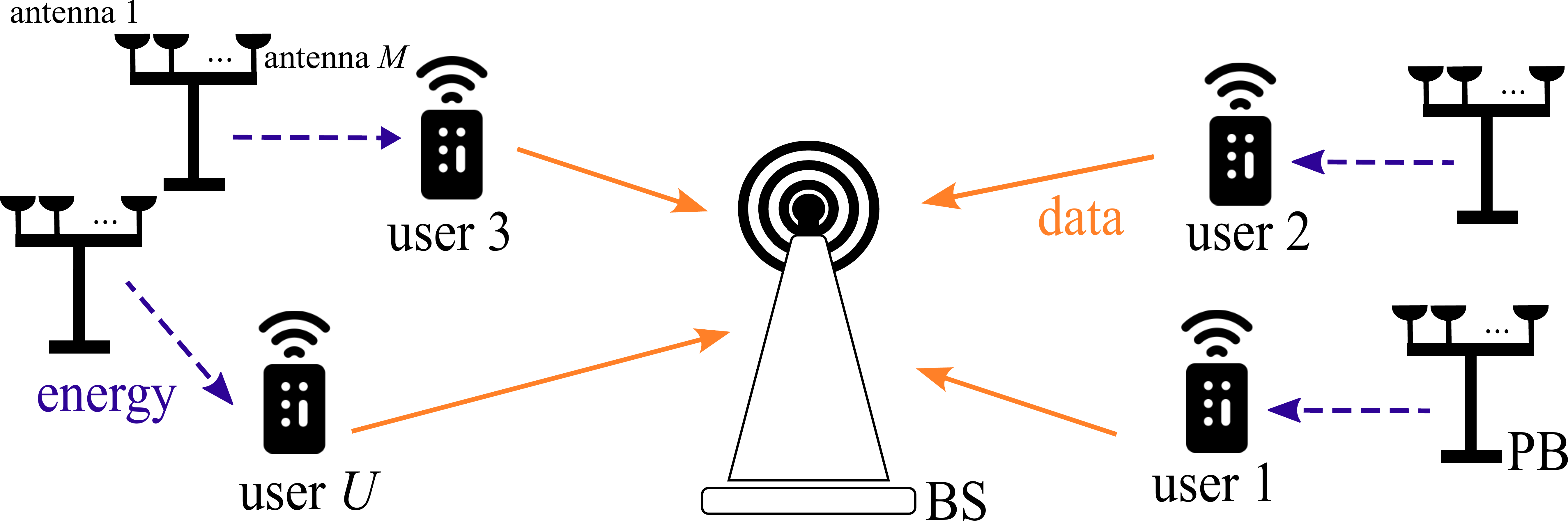}
  \caption{\Gls{wpt} with a \gls{bs} and $U$ user's \gls{rf}-\gls{eh} system model.}
\label{fig:system}
\end{figure}

We assume the scenario illustrated in Fig.~\ref{fig:system} where a set 
of $U$ single-antenna users are charged by personal (or own) \gls{pb}s. Each \gls{pb} is equipped with a \gls{ula} of $M$ antennas. The users need to harvest \gls{rf} energy from \gls{pb}s to send data to the \acrfull{bs}, which results in a waiting period referred to as the charging slot. The \gls{pb} transmits at a fixed transmit power during the charging slot. The user-\gls{bs} communications use \gls{irsa} channel access protocol where $K$ data slots $\{\mathrm{d}_k\}_{k=1}^{K}$ are allocated. Time is discrete and indexed by $t$. 

\subsection{\Acrfull{irsa}}
\begin{figure}[t]
  \centering
  \includegraphics[width=8.8cm]{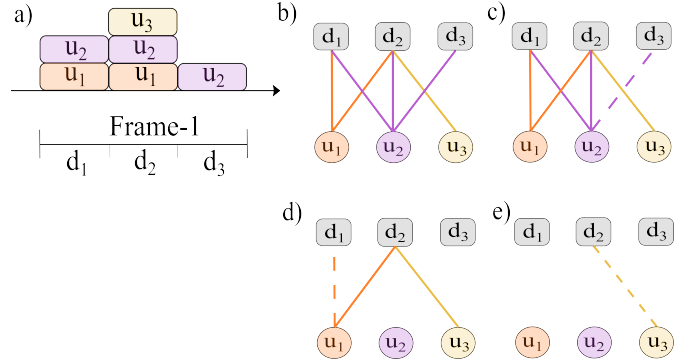}
  \caption{Bipartite representation of IRSA with its interference cancellation mechanism. (a) Frame structure. (b) Bipartite graph. (c) Iteration 1. (d) Iteration 2. (e) Iteration 3.}
\label{fig:irsalgo}
\end{figure}

In the context of \gls{irsa} for data transmission, a user employs multiple replicas of a packet within each frame as shown in Fig.~\ref{fig:irsalgo}a. These packets contain information about the slots in which other replicas are sent. The transmission of a packet and its replicas can be represented as a bipartite graph, as depicted in Fig.~\ref{fig:irsalgo}b. Each user is represented by a circle, while each data slot is represented by a square. The edges connect a user to the selected data slots. For an example, user $\mathrm{u}_1$ has chosen to transmit in slots $\mathrm{d}_1$ and $\mathrm{d}_2$, whereas user $\mathrm{u}_2$ transmits in slots $\mathrm{d}_1$, $\mathrm{d}_2$, and $\mathrm{d}_3$, while user $\mathrm{u}_3$ transmits only in slot $\mathrm{d}_2$.

Once the replicas are received, the \gls{bs}  decodes the packets starting from a data slot with no collisions (Fig.~\ref{fig:irsalgo}c) as follows~\cite{Ghanbarinejad.13}. The decoded packet serves as a reference to decode packets in other data slots in which, all copies of the decoded packet from the corresponding user are removed first. Then, the \gls{bs} seeks another collision-free data slot to decode the next packet (Fig.~\ref{fig:irsalgo}d). This is repeated (Fig.~\ref{fig:irsalgo}e) until all data from users are sequentially decoded.

We assume that user $\mathrm{u}$ sends $\psi \in \{0, \dots, N\}$ replicas of a given packet in time frame $t$ with probability $\pi_\psi$.  The probability mass function that determines the probability of sending a given number of replicas is represented as the polynomial $\vec{\pi}(x) = \sum_{\psi=0}^{N} \pi_\psi x^{\psi}$.

\subsection{Channel Model}
We assume users and the \gls{bs} are equipped with a single antenna and the channel between a user and the \gls{bs} (used for data transmission) is subject to Rayleigh fading. In addition, we assume a frequency hopping model, where transmission occurs in different channel realizations. Therefore, $g_u = \sqrt{1/2} \,\mathcal{CN}(\mathbf{0},PL)$ is modeled as a complex Gaussian random variable with zero mean and
$PL = 20\log_{10}{(4 c \pi d^{\text{BS}}_u/f)}$ variance where $c$ is the speed of light, $f$ is the carrier frequency, and $d^{\text{BS}}_{u}$ is the \gls{bs}-user distance.

In \gls{eh}, the channel between the \gls{pb} and each user is subjected to quasi-static Rician fading due to the \gls{los} connectivity. Thus, the channel vector between each user and the \gls{pb} is given by~\cite{goldsmith} 
\begin{equation}
    \mathbf{h}_{u} = \sqrt{\beta_{u}} \left(\overline{\mathbf{h}}_{u} + \tilde{\mathbf{h}}\right) \quad \in \mathbb{C}^{M \times 1},
\end{equation}
where $\beta_{u} = c^2/(16 \pi^2 f^2 (d^{\text{PB}}_{u})^{\alpha})$ is the average power gain,  $c$ is the speed of light, $f$ is the carrier frequency, $d^{\text{PB}}_{u}$ is the \gls{pb}-user distance, and $\alpha$ is the pathloss exponent. $\overline{\mathbf{h}}_{u} =  \sqrt{\kappa/(2\,(1 + \kappa))}  \, [1, e^{i \tau_1}, \cdots ,e^{i \tau_{M-1}}]^T$ is the deterministic \gls{los} component, and $\tilde{\mathbf{h}} \sim \sqrt{1/(1 + \kappa)} \mathcal{CN}(\mathbf{0},\mathbf{R})$ is the zero-mean scattering (random) component with covariance $\mathbf{R} = \EX\left[\tilde{\mathbf{h}} \tilde{\mathbf{h}}^H\right]$ where $\tau_m$, $m \in  \{1, \cdots ,M - 1\}$, is the mean phase shift of the $(m + 1)$-th array element with respect to the first antenna and $\kappa$ is the \gls{los} factor~\cite{Onel.21}. Assuming half-wavelength spacing between antenna elements $\tau_m = -m \, \pi \sin \theta$ is held, where $\theta \in [0,2\pi]$ is the azimuth angle relative to the boresight of the transmitting antenna array. 

\subsection{Incident Power}
Accounted for a single user, the \gls{mrt} precoding is used as the optimal precoder design. Hence, the exact channel $\mathbf{h}_{u}$ is used for the precoder with \gls{fcsi} while for \gls{acsi}, the expected channel $\EX[\mathbf{h}_{u}] = \sqrt{\beta_u}\, \overline{\mathbf{h}_{u}}$ is used. In this view, the harvested power is given by~\cite{Lo.99},
\begin{equation}
    P_U^i = 
    \begin{cases}
    \beta_{u} P_\mathrm{b} ||\mathbf{h}_{u}||^2, & i = \text{F-CSI}, \\
    \beta_{u} \, P_\mathrm{b} \left| ||\overline{\mathbf{h}_{u}}|| + \frac{\overline{\mathbf{h}_{u}}^H \tilde{\mathbf{h}}}{||\overline{\mathbf{h}_{u}}||} \right|^2 , & i=\text{A-CSI}.
    \end{cases}
\end{equation}

\subsection{Energy Model}
Assuming that user $u$ sends replicas $\psi_u^t$ in the time frame $t$, the energy level of user $u$ evolves as per
\begin{equation}\label{eq:ehfunc}
    E^{t}_{u} =  \min \left( E_0, \,E^{t-1}_{u} + G(P_{u})t_\mathrm{C} - \frac{ (1+\rho^{t}_{u})\xi_u |g_{u}|^2}{(d^{\text{BS}}_u)^\alpha} \right)
\end{equation}
where $\xi_u = t_\mathrm{T}P_{u}L$ is the energy spent in the transmission of one replica, $L$ is the packet size, $P_{u}$ is the transmission power, $t_\mathrm{C}$ is the charging slot time, $t_\mathrm{T}$ is the data slot time, and $G(P_{u}) = \mathcal{W}(1 - e^{-c_0 P_{u}})/(1+ e^{-c_0(P_{u} - c_1)})$ is a non-linear \gls{eh} function~\cite{Boshkovska, Clerckx.19}, with $\mathcal{W}$ being the saturation level, and $c_0$ and $c_1$ are constants. The battery capacity is $E_0 = \omega \xi_u$, corresponding to the energy required to send $\omega$ packets. We define energy levels as discrete. Note that, the maximum number of replicas user $u$ can send at time $t$ is $l_u^\mathrm{max} = \frac{E^{t}_{u}}{\xi_u} - 1$.

\section{Packet Decode Maximization Problem}
\label{sec:problem}

Our objective is to identify the optimal policy, denoted as $\phi$, that maximizes the successful packet decoding throughout a planning horizon of duration $H$. In this context, each user policy $\phi_u = \{\boldsymbol{o}_u^1, \boldsymbol{o}_u^2, \ldots, \boldsymbol{o}_u^H \}$ dictates the allocation of replicas to each user during individual time frames, where $\boldsymbol{o}_u^t$ is a vector of size $\psi$. We establish a collection of optimal policies denoted as $\Phi = \{\phi_1, \phi_2, \ldots, \phi_u\}$. The quantity of successfully decoded messages during time frame $t$ is represented as $\mathcal{R}_u^t$. Formally, our problem is formulated as follows:
\begin{subequations}
\begin{align}
\mathcal{R}_u =& \max_{\phi_u \in \Phi} \frac{1}{H} \EX \left[ \sum^{\infty}_{t=1} \mathcal{R}_u^t(\phi_u) \right]\label{eq:packdec}\\
 \text{s.t.} \qquad  
    \label{eq:c1}
	&1 \leq \psi  \leq N_\mathrm{max},\\
	\label{eq:c3}
	&0 \leq N \leq l_u^\mathrm{max},\\
	\label{eq:c4}
	&E_0 \leq E^{t}_{u}.
\end{align}
\end{subequations}

 We are particularly focused on achieving the highest possible success rate. However, we do not engage in separate power optimization or channel conditions. Instead, we maintain a constant transmit power setting that applies uniformly across all channel realizations. Additionally, we do not prescribe how users should utilize the energy they harvest; our primary concern is ensuring that each user receives sufficient energy to transmit a specified number of packet replicas. The exact number of replicas is contingent upon the energy available in each user, with a minimum requirement of at least one packet transmission per user. If a user has surplus energy, it has the flexibility to transmit additional replicas. Furthermore, we do not specify the precise number of replicas each user should send in each time frame but rather establish a maximum limit on the number of replicas that can be transmitted. Moreover, the allocation of packets to data slots is not predetermined; instead, it is determined randomly, avoiding fixed patterns.

 The problem present in~\eqref{eq:packdec} can be approached as a single-agent scenario, wherein a learning emergent protocol comes into consideration. In this setup, each individual agent strives to learn the most suitable method for transmitting replicas, all while taking their energy limitations and the quantity of sent packets into careful consideration. To tackle this problem effectively, we can employ the techniques of \gls{rl} as \textit{Q}-learning.

\section{\gls{rl} policy for packet replication}
\label{sec:RLsolve}

In \acrfull{rl}, a learning agent interacts with an environment to solve a sequential decision-making problem model as a discrete time \gls{mdp}. Formally, \gls{mdp} is defined as a tuple $(S, A, P, R)$. Here, $S$ is the set of all possible states, $A$ is the set of actions, $P$ is the transition probability function, and $R$ is a reward function. In the \gls{eh}-oriented data transmission scheme each user at time frame $t$ is an agent with state $s_u^t \in S_u$. 
We outline our \gls{mdp} as
\begin{enumerate}
    \item State Space $S_u$: A state is defined as the energy level of each user. At time frame $t$, state $s^t_u \in S_u$ corresponds to the amount of energy at user $u$. Thus, we have a discrete state space: $S = [0, \xi_u, 2\xi_u, \ldots, \omega \xi_u]$.
    \item Action Space $A_u$: An action is defined as the number of replicas sent by each user. For example, action means user $u$ sends $a^t_u$ replicas at time frame $t$. Furthermore, each user is able to send at most one packet per time frame. Consequently, the maximum number of packets each user sends is $N_u + 1$, and the maximum number of replicas each user sends is $l_u^\mathrm{max}$. Therefore, we have a discrete action space: $A_u = [0, 1, \ldots, N_u]$.
    \item Transition Probability $P$: The transition probability between states is unknown.
    \item Reward $R_u$: The reward is defined as the number of successfully decoded packets using \gls{irsa} channel protocol in \gls{pb} at each time frame as the reward. 
\end{enumerate}

\begin{algorithm}[]
\caption{Pseudocode for \textit{Q}-learning}
\begin{algorithmic} [1]
\STATE Initialize $\mu$, $\delta$ and $\textit{Q}(s_u, a_u)$ randomly
\FOR{each frame $t \in \tau$} 
\FOR{user $u \in U$} 
\STATE Observe the $s^t_u$  and generate $x \sim\ U[0,1]$
\IF{$x < \epsilon$}
\STATE Select an action randomly
\ELSE
\STATE Select an action: $a^t_u(s^t_u) = \amax_{a \in A} \textit{Q}(s^t_u, a)$
\ENDIF
\STATE Randomly select data slots for action $a^t_u$
\STATE Decode the packets using \gls{irsa}
\STATE Collect the reward $R^t_u$ and send to respective user
\STATE Update the Q-value $\textit{Q}(s^t_u, a^t_u)$ as show in~\eqref{eq:q-learning}
\ENDFOR
\ENDFOR
\end{algorithmic}\label{alg:q-learning}
\end{algorithm}

\subsection{\textit{Q}-Learning}
To solve the proposed \gls{mdp} problem we employ a \gls{rl} algorithm known as \textit{Q}-learning. It is a model-free algorithm, meaning it does not require prior knowledge of the underlying system dynamics. \textit{Q}-learning method determines the optimal policy that maximizes a given reward. \textit{Q}-learning uses a table, known as the \textit{Q}-table, to store action-value estimates for each state-action pair in the MDP~\cite{Sutton.99}. The action-value estimate, denoted by $\textit{Q}(s_u^t, a_u^t)$, represents the expected cumulative reward that an agent will receive by taking action $a_u^t$ in state $s_u^t$ and following a specific policy.

The \textit{Q}-learning algorithm relies on iteratively updating the \textit{Q}-values based on the observed rewards and the agent's experiences. At each time step, the agent selects an action based on an exploration-exploitation strategy, such as $\epsilon$-greedy, which balances between trying new actions and exploiting the current best-known actions. After taking an action, the agent receives a reward and transitions to a new state. The \textit{Q}-value for the previous state-action pair is then updated using Bellman's equation as follows~\cite{Sutton.99}:
\begin{equation}\label{eq:q-learning}
        \textit{Q}(s_u^t, a_u^t) = (1- \mu)\textit{Q}(s_u^t, a_u^t) + \mu(R_u^t+ \delta \max_{a \in A}(\textit{Q}(s_u^{t+1}, a_u))),
\end{equation}
where $\mu$ is the learning rate factor and $\delta$ is discount factor.

\subsection{Learning Algorithm}
We implement an independent \textit{Q}-learning algorithm where the users exchange information with the \gls{bs}, and each user employs \textit{Q}-learning to learn its own policy. We adopt $\epsilon$-greedy for action selection~\cite{Sutton.99}. Where with probability $(1 - \epsilon)$ the agent will select the highest \textit{Q}-value action. Otherwise, the agent will randomly select an action. To ensure convergence, we decay the $\epsilon$ value over time. Let $\tau$ be the total number of time frames. Algorithm~\ref{alg:q-learning} presents a standard \textit{Q}-learning algorithm for single agent \gls{rl}. First, it initializes the learning parameters and $\textit{Q}(s_u^t, a_u^t) $ randomly. Then, for each frame, $t$ each user $u$ observes the current energy state $s^t_u$ and generates a random number $x \in [0,1]$. Then to select an action we use $\epsilon$-greedy, where with probability $\epsilon$, where each user selects randomly an action. Otherwise, the user selects the action with the highest \textit{Q}-value. Based on the previously selected action $a^t_u$, each user randomly selects data slots for the action. Then, \gls{irsa} is applied to decode the packets and \gls{bs} collects the reward $R^t_u$. Then, user $u$ observes its next state $s^{t+1}_u$ and receives the reward from \gls{bs} to find the highest \textit{Q}-value for the new state. Finally, the users update the \textit{Q}-value.

\begin{table}[h!]
\centering
    \caption{Simulation Parameters}
    \label{tab:tparameters}
    \begin{tabular}{l|l}
    \hline
    \multicolumn{1}{c|}{\textbf{Description}} & \multicolumn{1}{c}{\textbf{Value}} \\ \hline
    \gls{eh} saturation level & $\mathcal{W} = 10.73$~mW \\ \hline
    \multirow{2}{*}{\gls{eh} unitless constants} & $c_0 = 0.2308$ \\
    & $c_1 = 5.365$ \\ \hline
    Distance between \gls{pb} and user & $d = 3$ m \\ \hline
    Distance between \gls{bs} and user & $H = 70$ m \\ \hline
    Number of antennas at the \gls{pb} & $M \in \{4, 8\}$ antennas \\ \hline
    Number of users & $u = 4$ \\ \hline
    \gls{los} parameters & $\kappa = 2$ dB \\ \hline
    Path-loss exponents & $\alpha = 2.7$ \\ \hline
    Carrier frequencies & $f = 2.5$ GHz \\ \hline
    Charging slot length  & $t_\mathrm{C} = 1$ ms \\ \hline
    Data slot length  & $t_\mathrm{D} = 1$ ms \\ \hline
    Maximum number of packets & $N = 5$ \\ \hline
    Packet size & $L = 21$ bytes \\ \hline
    Data Transmission Power & $P = 10$~mW\\ \hline
    Transmit Power of the \gls{pb} & $P_\mathrm{b} = 1$ W \\ \hline
    Learning rate factor & $\mu = 0.1$ \\ \hline
    Discount factor & $\delta = 0.1$ \\ \hline
    $\epsilon$-greedy value & $\epsilon = 0.5$ \\ 
    \hline
    \end{tabular}
\end{table}

\section{Numerical Results}
\label{sec:numerical}

\begin{figure}[t]
   \centering
   \includegraphics[width=8.8cm]{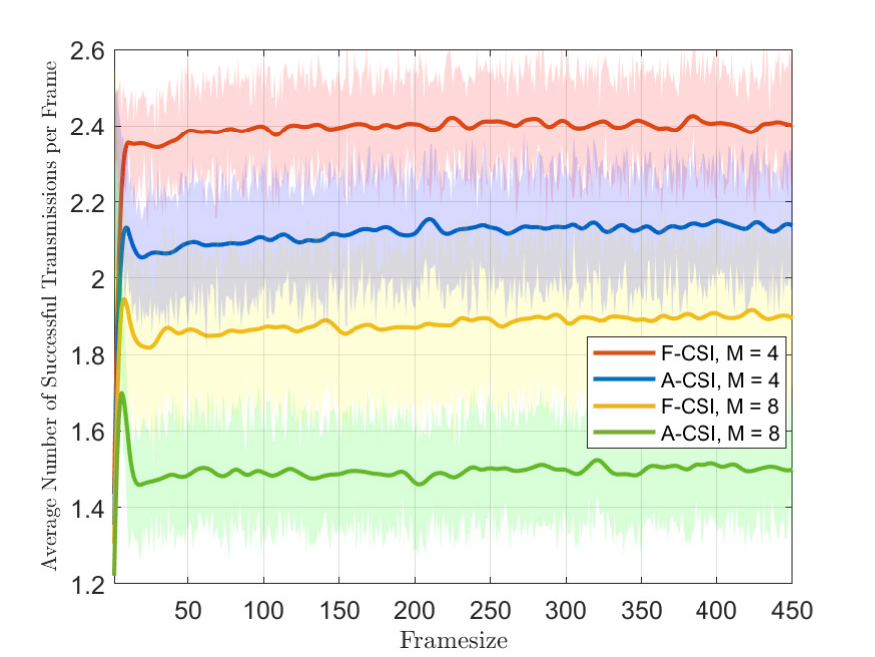}
   \caption{Convergence curve for \gls{fcsi} and \gls{acsi}.}
 \label{fig:CONV}
\end{figure}

In this section, we present numerical results to validate our scheme. We employ default simulation parameters, as enumerated in Table~\ref{tab:tparameters} unless specified otherwise. The propose \gls{rl}-based data transmission scheme that uses an \gls{irsa} protocol is compared with a baseline that uses an \gls{crdsa} data transmission scheme without learning~\cite{Meloni.12}. This baseline operates on the premise that the user always sends two replicas. However, if the user does not accumulate sufficient energy to transmit both replicas, only the main packet will be sent. We also compare the \gls{acsi} scheme, based on average channel estimation, and the \gls{fcsi} scheme, which assumes the availability of perfect instantaneous channel information for the user-\gls{pb} link.
It is important to mention that all the results presented here are based on averaging data gathered from ten simulation runs, with each run spanning $5000$ time frames. In Fig.~\ref{fig:CONV}, we observe the convergence behavior of the \textit{Q}-learning. We conduct $1000$ iterations, each comprising $450$ time frames. Notably, it becomes evident that a system with a greater number of antennas tends to converge quickly if compared with the same scheme with fewer antennas.

 \begin{figure}[t]
   \centering
   \includegraphics[width=8.5cm]{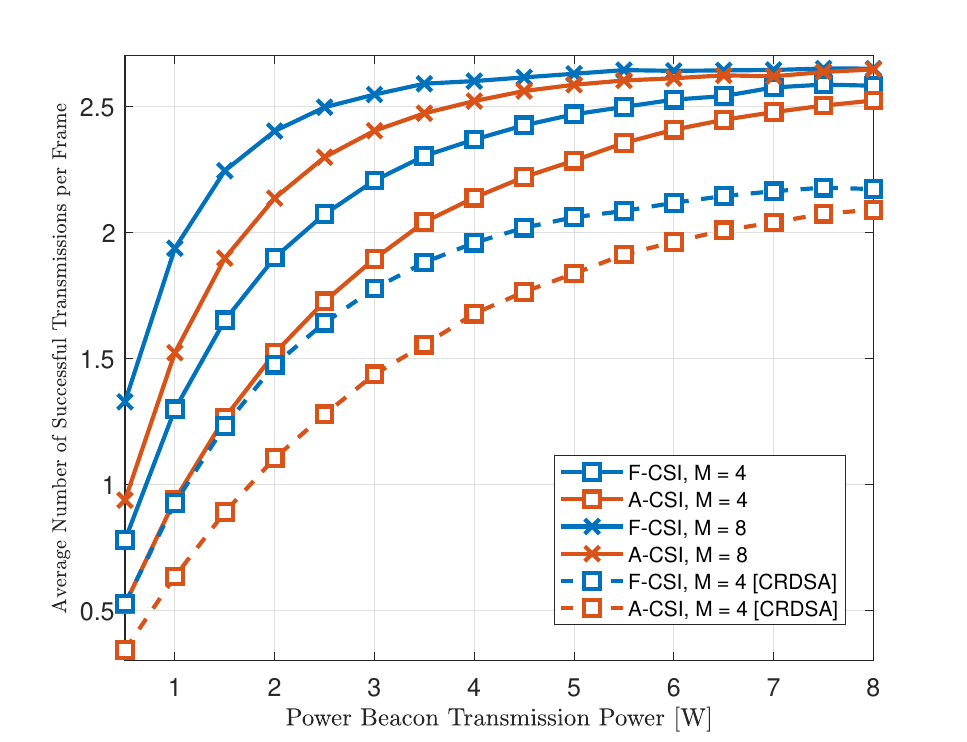}
   \caption{Impact of different \gls{pb} transmit power on the number of successful transmissions per frame.}
 \label{fig:POT}
 \end{figure}

 In Fig.~\ref{fig:POT}, we explore the relationship between the transmission power of the \gls{pb} signal and the average number of successful transmissions per frame. It is evident that the performance of all schemes experiences improvement as the PB transmission power is increased. As observed, we can affirm that the system with the learning will give $18\%$ more successful transmission if compared with the same scheme without the learning. Specifically, both \gls{fcsi} and \gls{acsi} schemes with $M = 8$ saturate at an average of $2.6$ successful transmissions per frame when the transmission power reaches $6$ W. Additionally, an interesting observation emerges as we scrutinize the gap between the \gls{acsi} and \gls{fcsi} schemes. This gap gradually diminishes as the \gls{pb} transmission power increases. For instance, when $P_\mathrm{b} = 2$ W, the \gls{fcsi} scheme achieves $19\%$ more successful packet transmissions compared to the \gls{acsi} scheme. However, this advantage decreases, and when $P_\mathrm{b} = 4$ W, the \gls{fcsi} scheme outperforms the \gls{acsi} scheme by $9\%$ in terms of successful transmissions.

   \begin{figure}[t]
   \centering
   \includegraphics[width=8.8cm]{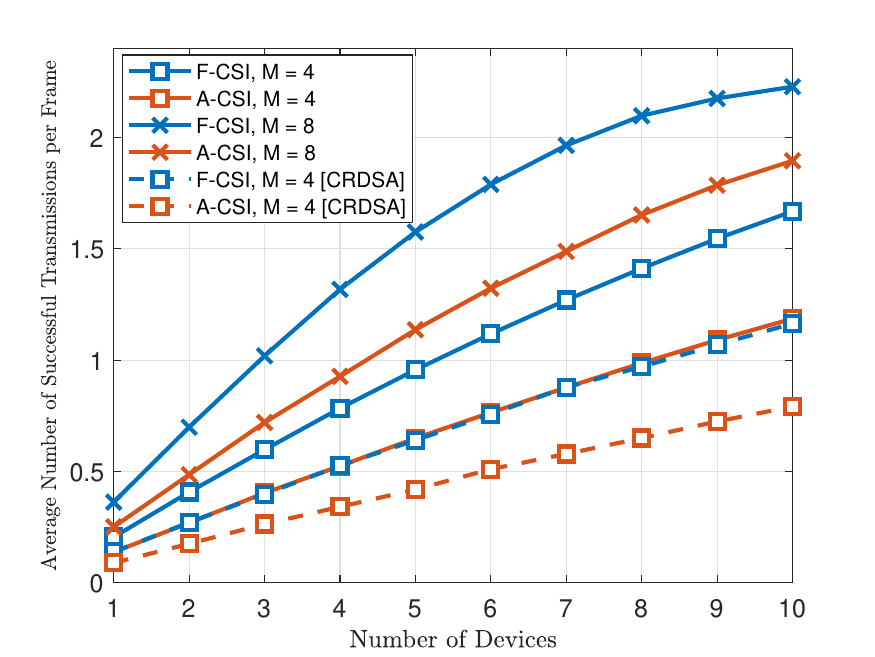}
   \caption{Impact of number of users on the number of successful transmissions per frame.}
 \label{fig:DEV}
\end{figure}

Figure \ref{fig:DEV} provides insight into how the number of users sharing information in the system impacts the number of successful transmissions per frame when the transmission power by \gls{pb} is $P_\mathrm{b} = 27$ dB. We observe an increase in the number of successful transmissions as the number of devices increases. Notably, when we compare \gls{acsi} with \textit{Q}-learning and a parameter $M=4$, its performance closely matches that of \gls{fcsi} \gls{crdsa} with $M=8$. This observation suggests that by using a simpler hardware structure and leveraging \textit{Q}-learning for average state estimation, it is possible to achieve an equivalent level of successful packet transmission.

\begin{figure}[t]
   \centering
   \includegraphics[width=8.8cm]{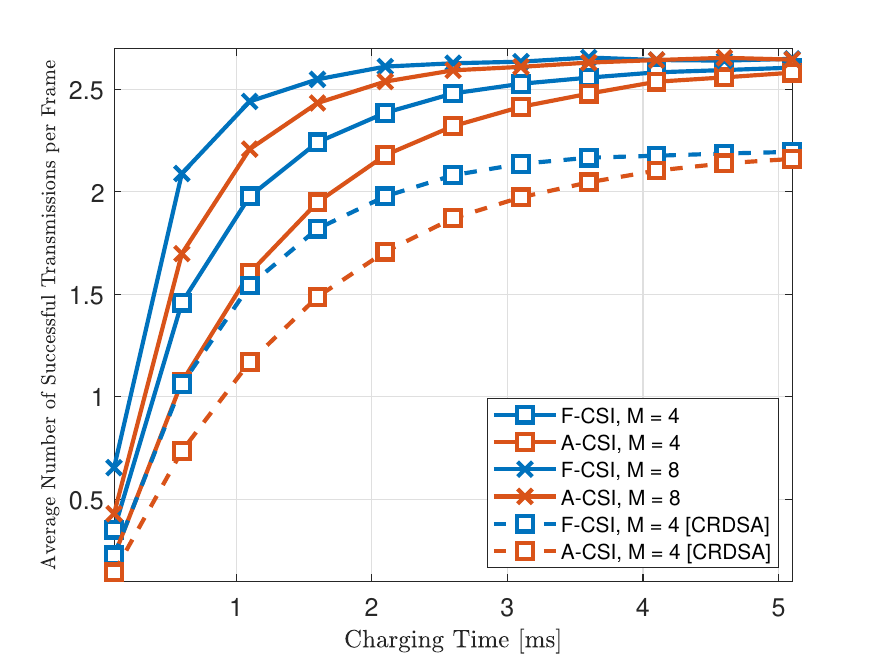}
   \caption{Impact of charging time on the number of successful transmissions per frame.}
 \label{fig:TIME}
\end{figure}

Next, in Fig.~\ref{fig:TIME}, we can see how the duration of charging time affects the number of successful transmissions per frame. It is important to note that the performance of all the different schemes improves as the charging time increases. Upon closer observation, we can identify a critical charging time that leads to the highest number of successful transmissions for each scheme. When comparing this system with the lack of a learning component, it achieves a notable $17\%$ increase in successful transmissions per frame compared to an equivalent system that utilizes the \gls{crdsa} channel protocol. For instance, in the case of \gls{fcsi} and \gls{acsi} schemes with $M = 8$, we achieve a total of $2.66$ successful transmissions when the charging time exceeds $4$ ms. This observation strongly supports the idea that there is a minimum required charging time to attain the optimal number of successful transmissions per frame.

\begin{figure}[t!]
   \centering
   \includegraphics[width=8.8cm]{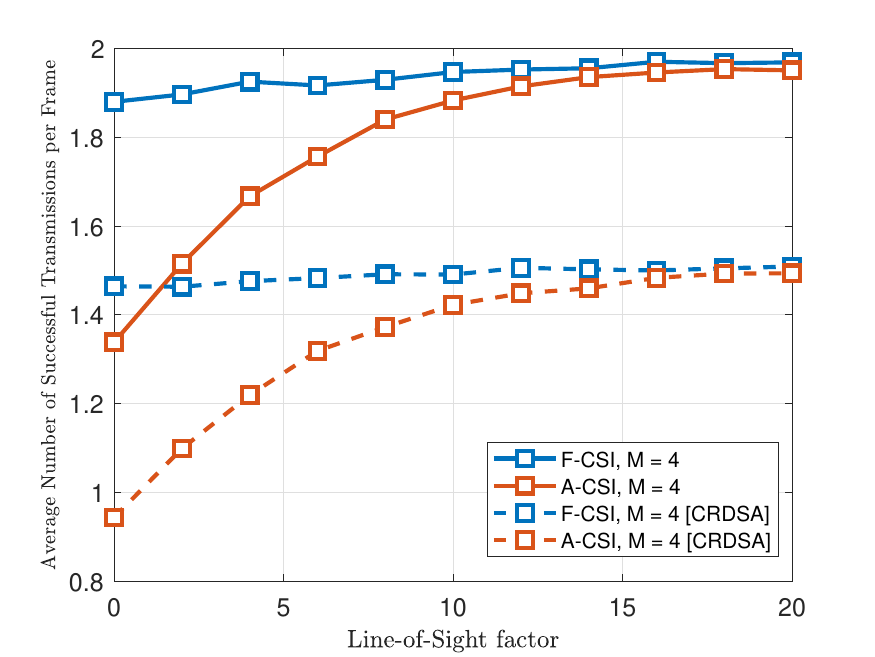}
   \caption{Impact of \gls{los} factor on the number of successful transmissions per frame.}
 \label{fig:LOS}
\end{figure}

Finally, in Fig.~\ref{fig:LOS}, we depict the \gls{los} factor's variation in relation to the average number of successful transmissions per frame. Notably, as we examine the graph, we observe a significant reduction in the gap between both schemes as the parameter $\kappa$ increases. In a scenario with severe fading conditions, such as when $\kappa = 1$ dB, we find that the \gls{fcsi} scheme with $M=4$ antennas achieves a $25\%$ higher number of successful transmissions compared to the \gls{acsi} scheme with an equivalent number of antennas. This performance gap diminishes to $9\%$ when $\kappa = 6$ dB. Additionally, it is worth noting that the \gls{fcsi} scheme, without the inclusion of action learning, outperforms a \gls{acsi} system with learning only when $\kappa$ is less than $2$ dB. These results underscore the efficacy of \gls{acsi} as a viable beamforming option, particularly when the \gls{pb} can be strategically positioned in a favorable configuration, enjoying a strong \gls{los} connection with the user nodes.

\section{Conclusion}
\label{sec:conclu}

This paper presents a \gls{eh}-oriented data transmission scheme where a set of single-antenna users is charged by their own \gls{pb} using \gls{acsi} beamforming. After the \gls{eh}, each user transmits the data to a \gls{bs} using the \gls{irsa} protocol. First, we characterize the closed-form expression for the energy model. The distributed \textit{Q}-learning algorithm finds the optimal policy that maximizes the number of successful transmissions per frame under energy constraints. As the numerical results show the \textit{Q}-learning method increases the number of successful transmissions per frame in $18\%$ if compared with the same scheme using \gls{crdsa} channel protocol without the learning process. Also, we can affirm that \gls{acsi} beamforming in the worst case achieved $19\%$ less successful transmission than the \gls{fcsi}. The optimal policy given by \textit{Q}-learning solution ensures a higher number of successful transmissions per frame for each user, than in the case of a transmission scheme not optimized with an \gls{rl} policy. 

\bibliographystyle{IEEEtran}
\bibliography{Qlearning}

\end{document}